# Realization of the anisotropic XY model in a Tb(III)-W(V) chain compound

Edoardo Pasca,[a] Tommaso Roscilde,[b] Marco Evangelisti,[c] Enrique Burzurí,[c] Fernando Luis,[c] L. Jos de Jongh,*[a] and Stefania Tanase*[d,e]

We present evidence of 1D $S=1/2$ anisotropic XY ferromagnetism in the paramagnetic phase of the cyanido-bridged chain complex $[Tb(pzam)_3(H_2O)W(CN)_8]\cdot H_2O$, with M=W(V), based upon the measurement of the specific heat, the uniform susceptibility and the magnetization curve on a powder sample. Both the specific heat and the susceptibility exhibit a transition peak to long-range ferromagnetism, mediated by the residual interchain couplings, at a critical temperature $T_c=1.15$ K. Yet in the temperature range $T > T_c$ the specific heat exhibits a broad Schottky anomaly which can be very well fitted by the exact result for the 1D $S=1/2$ XY model with strong anisotropy in the XY plane ($J_x = 1.89$ K, $J_y=2J_x$). For the analysis of the susceptibility and magnetization, not directly accessible theoretically via the exact solution of the XY model, we resort to numerically exact quantum Monte Carlo simulations. Very good agreement is found between the powder-sample data and the Monte Carlo data using an effective average g-factor.

## Introduction

Molecular magnetism enables assembling of magnetic nanostructures in a controlled way by synthetic means. This has led to a rich showcase of magnetic behaviours, ranging from superparamagnetism of single molecules[1] to collective phenomena.[2] The design of novel anisotropic and well-isolated one-dimensional (1D) magnetic systems is certainly among the challenging subjects. Although the experimental research on such systems started in the 1980s,[3] it has recently received a new impetus with the introduction of molecular materials.[4-6] The combination of ferro- or ferrimagnetic intra-chain coupling and slow dynamics, associated with the strong anisotropy, renders these so-called single-chain magnets (SCM) very appealing for the realization of e.g. magnetic memories at the atomic level.

In extended 1D polynuclear complexes, the presence of alternating $d$ and $4f$ metal centres is of special interest. This is because of the strong superexchange promoted by the former, combined with the peculiar crystal-field effects of the latter. Previously, we reported the synthesis and magnetothermal studies of heterometallic rare-earth (RE)-based cyanido-bridged 1D complexes, namely $[RE(pzam)_3(H_2O)Mo(CN)_8]\cdot H_2O$ (RE = Nd, Sm, Gd, Tb and Er), whose magnetic properties are mainly determined by the particular RE-ion involved.[7-11] We found that the Nd(III) and Mo(V) ions are coupled ferromagnetically into magnetic chains by an XY-type (planar) interaction.[10] The interaction between Tb(III) and Mo(V) is likewise ferromagnetic, but with a strong Ising-type anisotropy.[7] By contrast, the non-cancellation of the antiferromagnetically coupled Gd(III) and Mo(V) spins results in a ferrimagnetic Heisenberg chain.[8] An antiferromagnetic interaction is also observed between Sm(III) and Mo(V) as well as Er(III) and Mo(V), respectively.[9] The symmetry of the interaction is Ising-Heisenberg in case of Sm(III), whereas the compound with Er(III) behaves as a pure XY chain.[9] Although long-range order in pure 1D materials has to occur at $T = 0$ K, the presence of inter-chain interactions, albeit weak, shifts the ordering temperature to a finite value. As a matter of fact, the compounds containing Sm(III) and Gd(III) ions show a transition to 3D long-range magnetic order at $T_C \cong 0.6$ K and $T_C \cong 0.7$ K, respectively.[8-9] A similar magnetic transition is observed at $T_C \cong 1$ K for the Tb(III) analogue.[7] From the comparison of the (isostructural) complexes $[Gd(pzam)_3(H_2O)M(CN)_8]\cdot H_2O$ (M = Mo and W), it appeared that the strength of the isotropic exchange interaction increases by replacing Mo(V) with W(V), probably as a result of the larger radial extent of the magnetic $5d$ orbitals.[8]

[a] Dr. Edoardo Pasca, Prof. L. Jos de Jongh
    Kamerlingh Onnes Laboratory, Leiden Institute of Physics, Leiden University
    PO Box 9504, 2300 RA, Leiden, The Netherlands
    E-mail: jongh@physics.leidenuniv.nl
[b] Dr. Tommaso Roscilde
    Laboratoire de Physique, CNRS UMR 5672, Ecole Normale Supérieure de Lyon, Université de Lyon, 46 Allée d'Italie, Lyon F-69364, France
[c] Dr. Marco Evangelisti, Dr. Enrique Burzurí, Dr. Fernando Luis
    Instituto de Ciencia de Materiales de Aragón, CSIC – Universidad de Zaragoza, Departamento de Física de la Materia Condensada 50009 Zaragoza, Spain
[d] Dr. Stefania Tanase
    Leiden Institute of Chemistry, Gorlaeus Laboratories
    Leiden University
    PO Box 9502, 2300 RA Leiden, The Netherlands.
[e] Current address: Van't Hoff Institute for Molecular Sciences
    University of Amsterdam
    Science Park 904, 1098 XH Amsterdam, The Netherlands
    E-mail: s.grecea@uva.nl

Here we focus on the role of a similar replacement, but now in a system where the single-ion anisotropy associated with the rare-earth ion plays a central role. These are [Tb(pzam)$_3$(H$_2$O)M(CN)$_8$]·H$_2$O, where M = Mo and W, respectively (hereafter denoted as Tb-Mo and Tb-W). in both cases, their crystal structure is based on an infinite chain-like motif of alternating [Tb(pzam)$_3$(H$_2$O)]$^{3+}$ cations and [M(CN)$_8$]$^{3-}$ anions (Fig. 1). For the Tb-Mo compound we found that the spatially alternating Tb(III) and M ions both carry an effective spin $S = 1/2$, and their ferromagnetic coupling exhibits a strong Ising-type anisotropy (compare for instance the Ising nature of the coupling in Tb-Mo).[7] Here we show that, although the coupling in Tb-W is likewise anisotropic, its thermodynamic behaviour cannot be fitted by the predictions of an Ising Hamiltonian, but rather of a fundamental quantum variant thereof (the $S = 1/2$ $XY$ model with strong in-plane anisotropy). In one spatial dimension this model represents a cornerstone in the theory of quantum magnetism: it is an integrable quantum lattice model that can be solved exactly by mapping it to a one-dimensional gas of free fermionic quasiparticles.[12] Hence it exhibits unconventional (fractionalized) excitations with fermionic statistics.

In this paper we present evidence of 1D $S = 1/2$ anisotropic $XY$ ferromagnetism in the paramagnetic phase of the Tb-W compound based upon the measurement of the specific heat, the uniform susceptibility and the magnetization curve on a powder sample. Both the specific heat and the susceptibility show a transition peak to long-range ferromagnetism, mediated by the residual inter-chain couplings, at a critical temperature $T_c$ = 1.15 K. Yet, in the temperature range $T > T_c$ the specific heat exhibits a broad Schottky anomaly which can be very well fitted by the exact result for the 1D $S = 1/2$ $XY$ model with strong anisotropy in the $XY$ plane. Analysing susceptibility and magnetization is complicated by the powder nature of the sample, by the two non-equivalent $g$-tensors for the RE and M ions, and by the fact that these two thermodynamic quantities are not directly accessible via the exact solution of the $XY$ model. To solve this problem we applied numerically exact quantum Monte Carlo simulations, whose results show very good agreement with the experimental ones by using an effective uniform $g$-tensor as fitting parameter.

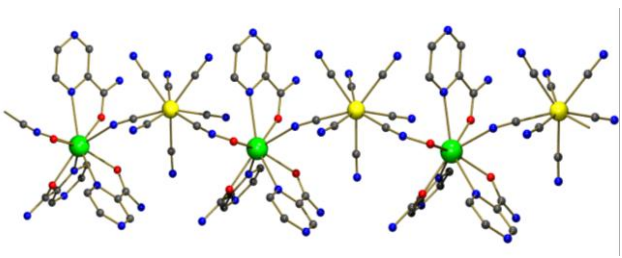

Figure 1. The 1D chain of [Tb(pzam)$_3$(H$_2$O)M(CN)$_8$]·H$_2$O, where M = Mo and W, viewed along the $b$ axis. Hydrogen atoms and the lattice water molecule are omitted for clarity. Tb, green; M, yellow; O, red; N, blue; C, grey.

## Results and Discussion

Figure 2 shows the temperature dependence of the magnetic susceptibility (as $\chi_M T$ versus $T$) of the Tb-W compound measured in the SQUID magnetometer at low fields in the range $T$ = 1.8 – 300 K. The room temperature $\chi_M T$ value of the Tb-W compound is 12.0 cm$^3$Kmol$^{-1}$, close to the expected value of 12.2 cm$^3$Kmol$^{-1}$ for non-interacting Tb(III) ($J_{Tb}$ = 6, $g_{Tb}$ = 3/2) and W(V) ($S_W$ = 1/2, $g_W$ = 2). This value remains almost constant down to ca. 150 K, below which it decreases, reaching the minimum of 10.2 cm$^3$Kmol$^{-1}$ at 12 K. At even lower temperatures, $\chi_M T$ increases markedly to 23.7 cm$^3$Kmol$^{-1}$ at 1.8 K. This behaviour is similar to that previously observed for the Tb-Mo analogue.[7] Accordingly, we assign the initial decrease of $\chi_M T$ to the depopulation of the Stark levels of the Tb(III) $^7F_6$ ground state, whilst the increase observed below ca. 12 K is ascribed to the ferromagnetic interaction between Tb(III) and W(V). Indeed, as shown in the inset of Figure 2, a plot of $\chi^{-1}$ versus $T$ in the temperature range below 10 K indicates a positive intercept. This implies a ferromagnetic interaction, with a Curie-Weiss temperature $\theta_W \cong$ 1.1 K. Relating this to the intra-chain exchange in the mean-field formalism yields the value 0.7 K for the exchange. A similar procedure applied to the Tb-Mo analogue yields the values $\theta_W \cong$ 0.7 K and 1.1 K for the exchange. The substitution of Mo(V) by W(V) thus leads indeed to an increase of the dominant interaction, as it happens for the isostructural Gd(III) analogues.[8]

Figure 2. Plot of $\chi_M T$ as a function of temperature for Tb-W. The inset shows a plot of $\chi^{-1}$ versus $T$ for Tb-W.

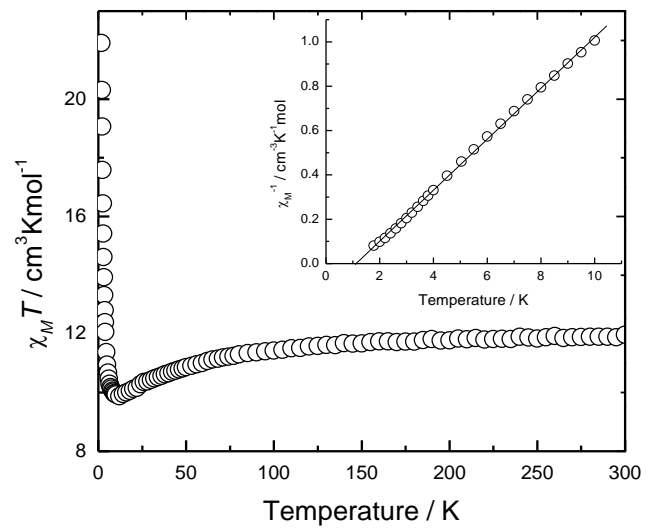

To confirm that the intra-chain interaction is indeed ferromagnetic, the field dependence of the magnetization was measured at low temperatures. Figure 3 shows the isotherm taken at 2 K. A field of $\approx$ 1 T is sufficient to saturate the magnetization of the Tb(III) and W(V) ground state magnetic moments. The subsequent slow and approximately linear magnetization increase up to 5 T reflects the contributions from the excited Tb(III) levels. Indeed, the slope of this high-field magnetization component yields a susceptibility value about equal to that measured in low field at around 50 K. Following the same procedure as for Tb-Mo,[7] we extrapolate the linear contribution to the magnetization in high field (above 2 T) down to zero field, obtaining from the intercept the value 3.6 $\mu_B$/molecule for the saturated ground state moment. Since the W(V) moment (with $g$ = 2.00 and $S$ = 1/2) is equal to 1 $\mu_B$ and the specific heat data to be discussed below evidence an effective spin 1/2 for Tb(III) in its ground state (below 10 K), we derive an effective powder $g$-value of $g_{eff}$ = 5.2 for Tb(III) in Tb-W, a bit lower than found previously for the Tb-Mo analogue ($g_{eff}$ = 5.8).[7]

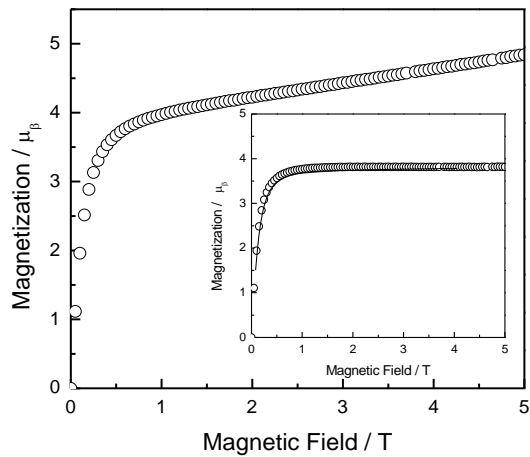

Figure 3. Field-dependence of the molar magnetization of Tb-W collected at $T$ = 2 K. The inset shows the magnetization of Tb-W with the contribution above 1 T from excited levels removed. Solid lines represent fit from the model described in the text.

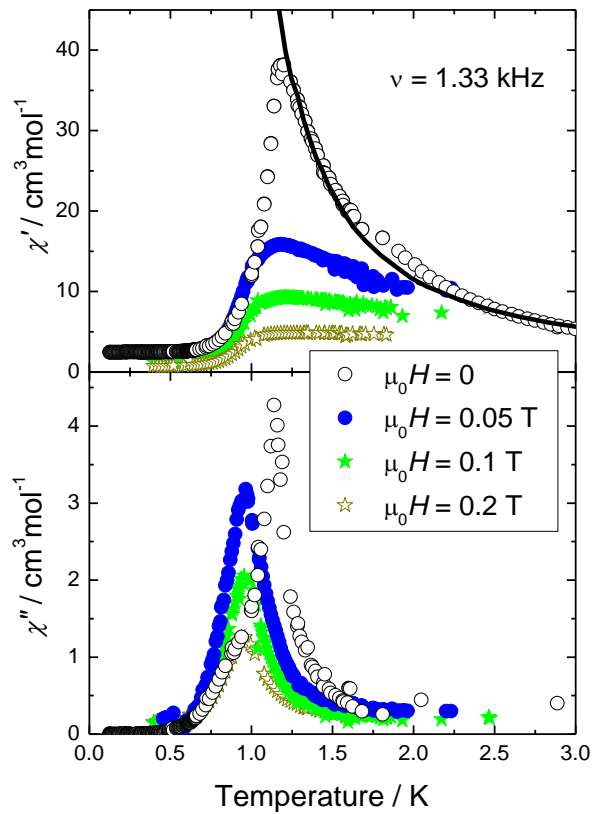

Figure 4. AC susceptibility measurements in different applied magnetic fields and at $\nu$ = 1.33 kHz for Tb-W. Upper and lower panels show the real and imaginary components, respectively. The solid line is the prediction from the model described in the text.

The susceptibility was further investigated by AC-susceptibility measurements down to 0.1 K. Figure 4 shows the real ($\chi'$) and imaginary ($\chi''$) components of the complex susceptibility. These are frequency independent in the range 100–1000 Hz; the $\chi'$ and $\chi''$ data for Tb-W plotted in Figure 4 are for 1033 Hz. The $\chi'$ at zero field shows a sharp peak at a temperature $T_C$ = 1.15 K, which is attributed to a transition to long-range 3D magnetic order induced by weak inter-chain couplings – most probably of dipolar origin. The divergence of the spin-correlation length along the magnetic chains triggers the establishment of 3D order below $T_C$. This involves a transition from 1D to 3D critical behaviour as $T_C$ is approached. Comparing with the analogous $\chi'$ data for the Tb-Mo compound,[7] we see a qualitative similarity. However, the height of the susceptibility peak is much lower than in Tb-Mo and the divergence is correspondingly less steep: it cannot be fitted by the exponential predicted by the Ising model. Another obvious though small difference is the slightly higher $T_C$ value of 1.15 K for Tb-W, as compared to 1.0 K for Tb-Mo. As the inter-chain interactions are very weak, i.e. of the order of 0.03 K as estimated for Tb-Mo,[7] one expects the transition to long-range 3D order to vanish when small magnetic fields are applied to the sample. This is confirmed by the in-field data shown in Figure 4, in particular by the rapid disappearance of the $\chi'$ peak at $T_C$. Although the $\chi''$ data suggest some remnants of an ordering to be present up to 0.2 T, we ascribe this to short-range ordered segments in the chains.

The presence of a magnetic phase transition in Tb-W is corroborated by specific heat ($C$) measurements. Figure 5 shows a $\lambda$-like peak at $T_C$ = 1.15 K in zero field data, which sits on top of a broader Schottky-like anomaly. When applying an external field, the $\lambda$-like peak quickly disappears and the magnetic energy levels are further split by the Zeeman interaction – accordingly the Schottky-like anomaly is shifted to higher temperatures with increasing applied field (top panel of Figure 5). The magnetic specific heat ($C_m$) is seen to be the predominant contribution below approximately 5 K. In order to separate this part from the nonmagnetic lattice (phonon) specific heat, which becomes predominant at higher temperatures, the latter was fitted to the well-known Debye $(T/\theta_D)^3$ term, i.e. the low-temperature limit expected for the phonon contribution (dashed line in Figure 5). The fit yields a Debye temperature of 54.4 K for Tb-W, as compared to 46.7 K for the Tb-Mo compound (N.B. it was an error in the published value for Tb-Mo).[7] After subtracting the lattice contribution, we obtain the zero-field $C_m$ of Tb-W, which is plotted in the bottom panel of Figure 5. From $C_m$, the magnetic entropy $S_m$ can be obtained by integration: $S_m(T) = \int_0^T (C_m(T)/T)dT$ (inset of Figure 5). This quantity is of particular interest since it provides a direct determination of the number of degrees of freedom, i.e. the number of relevant spin-levels. For both Tb-W and Tb-Mo, the so-obtained entropy content in the low temperature range agrees with the value of $2\ln(2S+1) = 1.38$ expected for two spins $S$ = 1/2. This is compelling evidence that the magnetic ground state of Tb(III) is an effective spin 1/2 for both materials. Finally, we note that $C_m$ of Tb-W shows a behaviour that is qualitatively similar to that of Tb-Mo,[7] but quantitatively reveals some essential differences. More specifically, the peak at $T_C$ is much less pronounced, and also the shape of the broad anomaly is different. In what follows we shall present the theoretical model that provides a satisfactory explanation for the behaviour observed for the Tb-W compound.

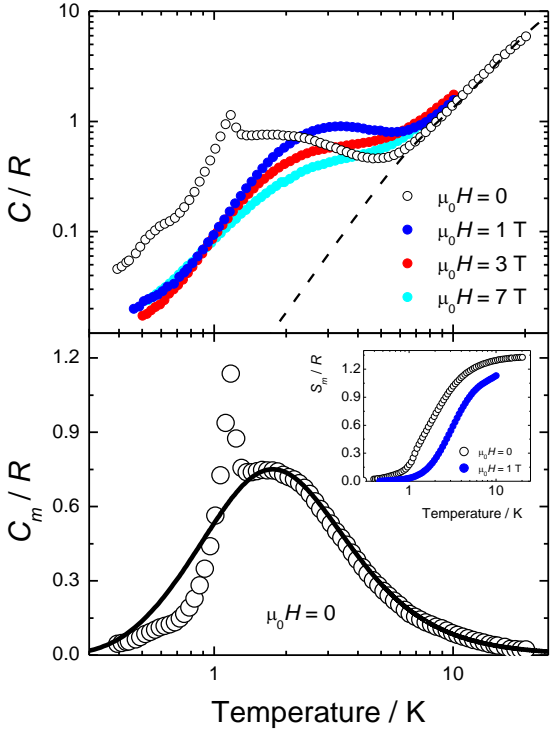

Figure 5. Top: Experimental specific heat of Tb-W, normalized to the gas constant $R$, collected for different applied fields up to 7 T, as labelled. The dashed line is the lattice phonon contribution. Bottom: Magnetic specific heat of Tb-W in zero applied field. The solid curve represents the model fit, described in the text. Inset: Magnetic entropy of Tb-W in zero field and in 1 T applied field. Discussion

As for Tb-Mo,[7] the low-temperature magnetic behaviour of Tb-W can be successfully modelled with a 1D spin Hamiltonian, with only nearest-neighbour interaction between the $d$-metal M ion (Mo or W) with the RE ion (Tb). In this approximation the interaction Hamiltonian can be written in the most general ($XYZ$) form as:

$$\hat{H} = -2 \sum_{i,\alpha=x,y,z}^{n} J_\alpha S_{i,RE}^\alpha (S_{i,M}^\alpha + S_{i-1,M}^\alpha) - \mu_B \sum_i^n \mathbf{H}(g_{RE}\mathbf{S}_{i,RE} + g_M \mathbf{S}_{i,M})$$

(1)

where the sum over $i$ is performed over the number $N$ of unit cells, noting that there is one RE and one M ion per unit cell. Since for the M ion we may safely assume an isotropic $g = 2.00$,[7-8, 10] the symmetry of the interaction with the magnetic field is determined by that of the $g$-tensor of the RE ion.

The previously reported magnetic behaviour of the Tb-Mo compound was successfully interpreted in terms of an assembly of parallel ferromagnetic chains, with an Ising-type interaction ($g_\parallel \gg g_\perp$) between Tb and Mo moments along the chains. The additional weak dipolar interactions between the chains are responsible for the transition to 3D long-range magnetic order at 1.0 K. For the Ising-type interaction of Tb-Mo we found $J_z = J_\parallel = 3.6$ K and $J_x = J_y = 0$.[7,13] For what regards the $g$-tensor literature values associated with the effective spin $S = 1/2$ of Tb are indeed often consistent with a strong, uniaxial Ising-type anisotropy, with $g_\parallel \approx 18$. For Tb-Mo a somewhat lower $g_\parallel = g_z \approx 10$, not uncommon for sites of low symmetry, was derived from the saturation moment, assuming $g_\perp = 0$.

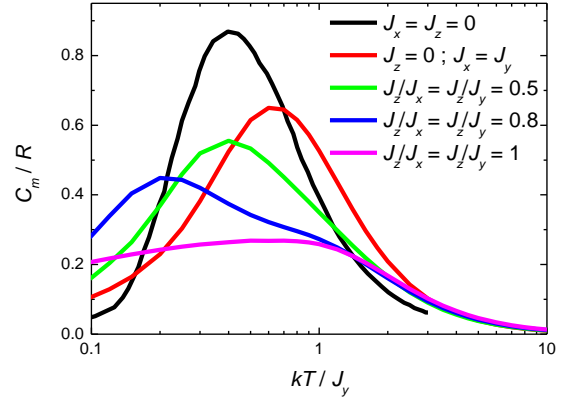

Figure 6. Predictions for the specific heat of the ferromagnetic chain model in the cross-over from Heisenberg ($J_x = J_y = J_z$) to XY ($J_x = J_y$; $J_z = 0$) models and in the pure Ising symmetry ($J_x = J_z = 0$; $J_y \neq 0$).[16]

In the case of Tb-W, when comparing the height of the observed Schottky-like specific heat maximum, *i.e.* $\approx 0.77$ (Figure 5), with theoretical predictions for ferromagnetic $S = 1/2$ chain systems with different interaction symmetries, it appears that its value is too low to be fitted with a pure Ising model ($J_x = J_z = 0$), whilst it is higher than any other model depicted in Figure 6. The only plausible model that could predict chain maxima comparable with our experiment appears to be the ferromagnetic $S = 1/2$ chain model with interaction intermediate between $XY$ and Ising. The corresponding anisotropic $XY$ interaction Hamiltonian can then be written as:

$$\hat{H} = -2J \sum_i [S_{i,RE}^x(S_{i,M}^x + S_{i-1,M}^x) + \lambda S_{i,RE}^y(S_{i,M}^y + S_{i-1,M}^y) - h(g_{RE}\mathbf{S}_{i,RE} + g_M \mathbf{S}_{i,M})]$$

(2)

where $h = g\mu_B H/(2J)$. Fitting $C_m$ of Tb-W with the specific heat calculated using eq. 2 gives $\lambda \approx 2$, $J = 1.89$ K and the solid curve depicted in the bottom panel of Figure 5. This model has the particularly appealing property of being exactly solvable, because it can be reduced to a model of non-interacting fermions via the Jordan-Wigner transformation. The non-interacting fermions have the following spectrum:

$$\epsilon_k = \pm 2J\sqrt{1-(1-\gamma^2)sin^2 k}$$

(3)

Here $\gamma = (1-\lambda)/(1+\lambda)$. The specific heat can therefore be calculated easily as the one of a gas of one-dimensional fermions with zero chemical potential. On basis of the $J$-values, we can calculate the components of the $g$-tensor for Tb(III) in Tb-W by means of the relation $J_y/J_x \approx (g_y/g_x)^2$, together with the formula $g_{powder}^2 = 1/3(g_x^2 + g_y^2 + g_z^2)$ for the powder $g$-value, which was estimated to be 5.2 from the saturation moment. This leads then to $g_x = 5.2$, $g_y = 7.4$, and $g_z = 0$. Apparently, the substitution of Mo(V) by W(V) in this crystal lattice leads to a drastic change in the symmetry of the *g*-tensor (in spite of the fact that the powder *g*-values are not very different).

The calculation of the susceptibility within this model is non-trivial because, in the fermionic language, the spin-spin correlations functions are non-local objects containing a string of operators on all the sites connecting the two sites of interest. For the calculation of the susceptibility we resort therefore to a quantum Monte Carlo method that is an original generalization of the Stochastic Series Expansion approach with directed loop algorithm,[14] valid for models having axial symmetry only.[15] From a theoretical point of view, the quantity of interest is the (molar) susceptibility along the α (= X, Y, Z) axis, which, assuming that the *g*-tensor is diagonal, takes the form:

$$\chi^{\alpha\alpha}(T) = \frac{\partial^2 F}{\partial (H^\alpha)^2} =$$
$$\frac{\mu_B^2 N_{av}}{JN_s} \sum_{ij} \sum_{X,X'=M,RE} g_X^{\alpha\alpha} g_{X'}^{\alpha\alpha} \int_0^{2\beta J} d\tau \langle S_i^\alpha(0) S_j^\alpha(\tau) \rangle \chi^{(p)} \quad (4)$$

where $\beta = 1/kT$, *F* is the (molar) free energy, $N_{av}$ is the Avogadro number while $N_s$ is the size of the spins in the simulation box. Given that are two spins per unit cell, we consider 2 moles of spins per mole of compound.

In practice the experimental susceptibility measured on powder samples can be thought of as the angle average of the susceptibility measured along all possible axes:

$$\chi^{(p)}(T) \sim \frac{1}{4\pi} \int d\varphi d\theta \sin\theta \left( \sum_{ij} \int_0^{2\beta J} d\tau \left\langle S_i^{(\theta\varphi)}(0) S_j^{(\theta\varphi)}(\tau) \right\rangle \right) \quad (5)$$

where $S^{(\theta\phi)} = \sin\theta\cos\phi S^x + \sin\theta\sin\phi S^y + \cos\theta S^z$. It is relatively straightforward to show that for the anisotropic *XY* model in zero applied field, $\langle S_i^\alpha(0) S_j^\beta(\tau) \rangle = 0$ for all $\alpha \neq \beta$. Hence the powder susceptibility reduces to:

$$\chi^{(p)}(T) = \frac{1}{3}(\chi^{xx} + \chi^{yy} + \chi^{zz}) \quad (6)$$

Not knowing a priori the *g*-tensors of the two different ions, we take the drastic assumption of fitting the experimental data (Figure 4) to Eq. (4) and (6) assuming a single ion per unit cell, and with an effective *g*-tensor proportional to the identity, $g_{eff}$. We found $g_{eff} \sim 2.24$, which can be thought of as an average *g* factor for all spin components of both magnetic ions.

Similarly, the theoretical expression for the powder magnetization in an applied field can be taken as the angle average of the magnetization (per unit cell) of the XY model in an externally applied field:

$$M^{(p)}(H;T) = \frac{\mu_B}{4\pi} \frac{2}{N_s} \sum_{i,X} \int d\theta d\varphi \sin\theta \langle (g_X \mathbf{S}_{i,X})^{(\theta\varphi)} \rangle_{H^{(\theta\varphi)}} \quad (7)$$

where $\langle S_i^{(\theta\phi)} \rangle_{h^{\theta\phi}}$ is the statistical average of the spin XY model in a field applied along the (θφ) direction.

Assuming that the g-tensor is diagonal, we have that:

$$(g\mathbf{S})^{(\theta\phi)} = g^{xx} \sin\theta \cos\varphi S^x + g^{yy} \sin\theta \sin\varphi S^y + g^{zz} \cos\theta S^z \quad (8)$$

Using Eq (7) would imply to fit the experimental data using angle averaged theoretical data with 6 parameters (3 *g*-tensor components per ion), after having calculated the response of the system to a field oriented along all possible spatial directions and having averaged over the angles.

Given that *y* is the easy axis of the system, and due to the large anisotropy between the y axis and the other two, we can assume that the powder magnetization is dominated by the response of each chain along its local y axes. Therefore, to compare with reasonably simple theoretical data, we assume that the magnetization of the powder sample can be approximated by the response of the XY chain immersed in an "average field" along the y axis. The theoretical calculation of the magnetization of the XY model along the easy (y) axis requires the use of QMC, because the XY model in a longitudinal field is not an exactly solvable model.

Even having at our disposal the QMC data for the XY chain in a longitudinal field, we are still faced with the problem that the RE and M ion actually see different effective magnetic fields, due to their different g factors. Hence we should need to calculate the magnetization curve of the XY model in a longitudinal field which is spatially periodic (with period 2), leaving the difference in field amplitude from the RE sites to the M sites as a fitting parameter. This requires again calculations that are beyond the scope of the present analysis. We therefore approximated that both the RE and the M site see the same effective average field $h = \tilde{g}\mu_B H/(2J)$, and that the RE ion develops a magnetic moment $\tilde{g}\mu_B S$, while the M ion develops a magnetic moment $2\mu_B S$ with gyromagnetic factor 2. This then give us the fitting formula for the magnetization per unit cell:

$$\frac{M_{fit}(T,H)}{\mu_B} = (\tilde{g} + 2) \, m_{th}^y(\tilde{g}\mu_B H/2J) \quad (9)$$

Where $m_{th}^y(\tilde{h})$ is the easy axis magnetization of the XY model in a uniform field *h*.

We get an excellent agreement between theory and experiment for $\tilde{g} = 5.2$, which is imposed by the saturation value of the magnetization after eliminating the linear drift at high fields shown by the experimental data (Figure 3) - this value turns out to be the correct factor for the rescaling of the field axis.

## Conclusions

Our studies show that the substitution of Mo(V) by W(V) in the series [Tb(pzam)$_3$(H$_2$O)M(CN)$_8$]·H$_2$O (M = Mo or W) gives a drastic change in the symmetry of the Tb(III) *g*-tensor. Although the powder *g*-value is almost the same, the underlying anisotropy of the *g*-tensor is very different. The magnetic interaction between the Tb(III) and M(V) changes from Ising-type into an anisotropic *XY* form. Here, we presented a detailed analysis of the magnetic data of the Tb-W compound in terms of theoretical predictions for the 1D *XYZ* Hamiltonian.

## Experimental Section

The synthesis and complete crystallographic studies of the family of complexes having general formula [RE(pzam)$_3$(H$_2$O)M(CN)$_8$]·H$_2$O are reported elsewhere. [7-8, 10] They crystallize in the space group *Pna*21 and their crystal structure is formed by chains of cyanido-bridged alternating arrays of [Mo(CN)$_8$]$^{3-}$ and [RE(pzam)$_3$(H$_2$O)]$^{3+}$ fragments running along the *b* crystallographic axis (Fig. 1). Six equivalent chains are interacting through hydrogen bonding interactions, giving rise to a 3-dimensional network of weakly coupled chains.

Spectroscopic and X-ray powder crystallographic studies have shown that the analogue compounds with W(V) are isostructural.

The thermal and magnetic properties of the two compounds were measured at low temperature, in zero field and in applied magnetic fields, using both commercial and home-built equipment. Magnetization and susceptibility down to 1.8 K were measured in a Quantum Design MPMS, equipped with a maximum field of 5 T. In the temperature range 0.1 K – 2 K a home-made AC susceptometer housed in a dilution refrigerator was employed. The samples were in the form of powders and were mixed with a few percent of Apiezon-N grease to enhance the internal thermal contact. Data were corrected for the magnetization of the sample holder and for diamagnetic contributions as estimated from the Pascal constants.

Specific heat was measured in a Quantum Design PPMS, equipped with a $^3$He refrigerator, capable to measure from 300 down to 0.3 K, and in a home-built calorimeter housed in a dilution refrigerator, fit to measure in the range 0.1 K – 8 K. Both set-ups were equipped with a superconducting magnet, providing fields up to 7 T. The thermal relaxation method was used in both cases, with time constants ranging from 1 to 100 s in the home-made low-temperature calorimeter. Also these samples were in the form of micro-crystalline powder and were mixed with Apiezon-N grease to provide good internal thermal contact between crystallites, heater and thermometer. Typically, about 70% of the sample-grease mixture consisted of crystallites. A careful calibration of the addenda was performed separately.

## Acknowledgements

*This research was supported by a Veni grant from the Netherlands Organization for Scientific Research (NWO) to S. T. M. E., F. L. and E. B. acknowledge grants MAT2009-13977-C03 and CSD 2007-00010.*

**Keywords:** ((cyanide · terbium · magnetic · anisotropy · modelling))